\documentclass[runningheads,a4paper]{llncs}

\usepackage{amssymb}
\usepackage{graphicx}

\usepackage{url}
\urldef{\mailsa}\path|{brittka, humblets}@ornl.gov| 
\newcommand{\keywords}[1]{\par\addvspace\baselineskip
\noindent\keywordname\enspace\ignorespaces#1}

\usepackage{color}

\begin{document}

\mainmatter  % start of an individual contribution

\title{Instruction Set Architectures\\for Quantum Processing Units}

\titlerunning{ISAs for QPUs}

\author{Keith A. Britt%
\thanks{This contribution has been authored by UT-Battelle, LLC, under Contract No. DE-AC0500OR22725 with the U.S. Department of Energy. The United States Government retains and the publisher, by accepting the article for publication, acknowledges that the United States Government retains a non-exclusive, paid-up, irrevocable, world-wide license to publish or reproduce the published form of this manuscript, or allow others to do so, for the United States Government purposes. The Department of Energy will provide public access to these results of federally sponsored research in accordance with the DOE Public Access Plan.}%
\and Travis S. Humble}
\authorrunning{K. A. Britt and T. S. Humble}

\institute{Quantum Computing Institute\\
Oak Ridge National Laboratory\\
Oak Ridge, Tennessee USA 37830\\
\mailsa\\
\url{http://quantum.ornl.gov}}


\toctitle{Lecture Notes in Computer Science}
\tocauthor{Authors' Instructions}
\maketitle

\begin{abstract}
Progress in quantum computing hardware raises questions about how these devices can be controlled, programmed, and integrated with existing computational workflows. We briefly describe several prominent quantum computational models, their associated quantum processing units (QPUs), and the adoption of these devices as accelerators within high-performance computing systems. Emphasizing the interface to the QPU, we analyze instruction set architectures based on reduced and complex instruction sets, i.e., RISC and CISC architectures. We clarify the role of conventional constraints on memory addressing and instruction widths within the quantum computing context. Finally, we examine existing quantum computing platforms, including the D-Wave 2000Q and IBM Quantum Experience, within the context of future ISA development and HPC needs. 
\end{abstract}
%%%%%%%%%%%%%%%%%%%%%%%%%%%%%%%%%%%%%%%%%%%%%%%%%%%%%%%%%%%%%%%%%%%%%%%%%%%
\section{Quantum Processing Units}
The realization of quantum processing units (QPUs) represents a milestone in computing. For decades theoretical computational complexity gains using QPUs have served as a lure to solving conventionally intractable problems. As an example, using two different models of quantum computing Grover's quantum search algorithm finds a marked item in an unordered database of size $N$ in $O(\sqrt{N})$ whereas the best classical approach, a sequential search, requires $O(N)$ \cite{Grover1996,Hen2016}.
\par
QPUs harness these gains in algorithmic efficiency by preparing quantum physical systems using superposition and entanglement. Superposition is a state of a quantum mechanical particle storing mutually orthogonal values simultaneously within a single physical degree of freedom, e.g., position, spin, etc. Entanglement is the feature that the joint state of multiple particles may be correlated even in the absence of a physical communication channel between them. Using superposition and entanglement, QPUs initialize, store, and process quantum bits of information, i.e., \textit{qubits}, by manipulating registers of quantum physical systems.
\par
The gate model (or circuit model) of quantum computation closely matches the discrete set of operations found in conventional computing models. Sequences of one- and two-qubit gates represent the fundamental logic for transforming a quantum state. However, there are several unique features for quantum computing including the inability to copy, or clone, arbitrary quantum states. Thus the number of inputs into the circuit must be equivalent to the number of outputs \cite{Nielsen2011}.
\par
\begin{figure}
\centering
\includegraphics[width=\textwidth]{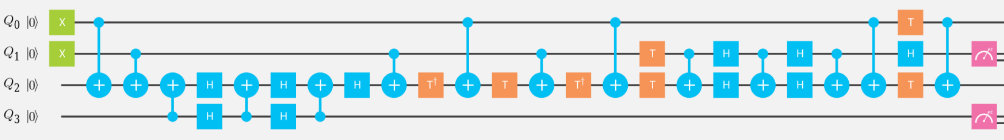}
\caption{This quantum circuit diagram describes the implementation of a half-adder using the gate model. This implementation was made using the QPU available from the IBM Quantum Experience \cite{IBM2017}. The registers $Q_0$ and $Q_1$ are initialized to $1$ using the $X$ gate. Due to hardware constraints, not all qubits can serve as the target of a CNOT operation. Therefore, several $swap$ operations are performed to facilitate interactions between qubits. The results of the half-adder, $Sum$ and $Carry$, are output on qubits $Q_3$ and $Q_1$, respectively.}
\label{fig:ibm}
\end{figure}

Several small-scale QPUs based on gate model designs have been demonstrated and a few are available for use outside of laboratory settings. Notably, the IBM Quantum Experience is accessible via the internet and allows users to construct a circuit using 5 qubits and up to 80 gates per qubit  \cite{IBM2017}.  This capacity for a QPU is still only useful in verifying not-to-scale quantum algorithms and empirical analysis of the reliability of the physical components. Interesting toy problems, like the half-adder depicted in Fig.~\ref{fig:ibm} are also possible.
%%%%%%%%%%%%%%%%%%%%%%%%%%%%%%%%%%%%%%%%%%%%%%%%%%%%%%%%%%%%%%%%%%%%%%%%%%%
%%%%%%%%%%%%%%%%%%%%%%%%%%%%%%%%%%%%%%%%%%%%%%%%%%%%%%%%%%%%%%%%%%%%%%%%%%%
\par
Another model of quantum computation is adiabatic quantum computation (AQC), which operates without discrete operations and no sequential constraints on the algorithmic steps \cite{Farhi2000}. In contrast to a discrete sequences of gates, the AQC model uses a continuous-time process during which the energetic interactions between register elements changes. Provided these changes are sufficiently slow, i.e., adiabatic, the evolution of the register can be well-defined relative to its energy eigenstates. In particular, if the register is initialized in an energetic ground state of the system Hamiltonian, it will remain in the energetic ground state under adiabatic evolution. The adiabatic model of computation is equivalent to the gate model in terms of computational power and the set of problems which can be efficiently solved. This equivalence arises through the influence of energetic changes on the computational state. Because the computational state is encoded by the energy eigenstates, changing these states is equivalent to performing a logical operation. The fidelity of this operation is controlled through the duration over which the energy changes.
\par
\begin{figure}
\centering
\includegraphics[width=3.5cm]{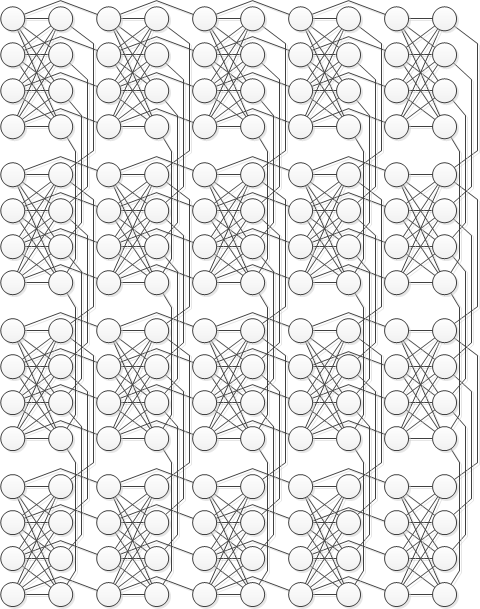}
\caption{The D-Wave Chimera graph in which each node represents a qubit and each edge represents a coupler. The D-Wave 2000Q, contains approximately 2000 qubits, expanding the grid of $K_{4,4}$ unit cells from 4 x 4 as shown here to 16 x 16.}
\label{fig:chimera}
\end{figure}

AQC devices are also currently available from D-Wave Systems. These quantum annealing devices implement a selected subset of the AQC model that restricts the available problems to finding the energetic ground state of an Ising Hamiltonian. Notably, the Ising problem is NP-Hard in general and also equivalent to a Quantum Unconstrained Binary Optimization (QUBO) problem. The latest D-Wave 2000Q hardware system is composed of approximately 2000 physical qubits
%can implement approximately 129 fully connected logical qubits across approximately 2000 physical qubits 
arranged in a topology called Chimera (see Fig.~\ref{fig:chimera}) \cite{DWave2017,Hamilton2016}. This is a sufficient number of qubits to enable direct comparison with modest-sized domain-specific problems. There has yet to be any demonstration that the D-Wave hardware can outperform a best-in-class classical computing system for any particular problem type, but there have been demonstrations where the D-Wave outperforms with respect to specific problem instances. As a means to address the probability of erroneous computations, the current D-Wave QPU relies on statistical sampling over repeated runs of identical programs to boost the confidence of observing the correct solution.
%%%%%%%%%%%%%%%%%%%%%%%%%%%%%%%%%%%%%%%%%%%%%%%%%%%%%%%%%%%%%%%%%%%%%%%%%%%
%%%%%%%%%%%%%%%%%%%%%%%%%%%%%%%%%%%%%%%%%%%%%%%%%%%%%%%%%%%%%%%%%%%%%%%%%%%
\par
Practical quantum processing is expected to require tens of thousands of qubits or more to solve problems with real world implications. The problem sizes for which quantum systems will surpass conventional computing system is still largely unknown due to the influence of engineering constraints on QPU performance. Nevertheless, a large QPU capacity is expected to be necessary. This requirement is underscored by additional technological realities. First, the degree of connectivity for any qubit is likely to be limited due to hardware fabrication constraints. This leads to the concept of virtualizing logical qubits using many physical qubits. Second, QPUs require very low noise environments to suppress erroneous computation. This typically implies working at cryogenic temperatures (below $1 K$), but even then quantum registers are vulnerable to errors. Quantum error correction protocols are necessary to reach fault-tolerant operation, and this incurs a substantial overhead for redundantly encoding a logical qubit using many (virtualized) physical qubits. Finally, the large problem sizes at which performance crossovers occur are very likely to require a large number of intermediate variables in which the numerical representation of each variable requires many logical qubits.
\par
Despite their potential, QPUs are not expected to act as stand alone devices but rather to require interactions with conventional central processing units (CPUs) and memory. Even though universal quantum computation is theoretically possible, near-term QPUs are most likely to be used as special-purpose processors. For example, the use of QPUs as accelerators for high performance computing (HPC) systems has already begun to mimic recent efforts with graphical processing units (GPUs) and today's leading scientific computing systems \cite{Top5002016,Britt2015,Fu2016}.
\par
To date, the CPU-QPU interface for gate-model processors has largely focused on quantum assembly (QASM) instructions. Originally introduced as a visual modeling language, QASM operates as a sequential, ordered list of discrete instructions representing gates acting on qubits \cite{Chuang2004}. This language naturally enforces the view that quantum instructions are pre-processed and queued for execution on the quantum hardware, with all the operations in a given time slice being executed simultaneously, e.g., within the same clock cycle. Similarly, the special-purpose AQC device from D-Wave defines interactions between the CPU and QPU via quantum machine instructions (QMI), a language that describes the continuous-time change in state of the processor Hamiltonian. This means the device settings for all the interactions between qubits are applied and executed concurrently.
\par
QASM and QMI are currently accepted by the quantum computing community as convenient methods for programming small-scale QPUs. Yet both approaches are unlikely to scale to larger processor sizes due to growth in the number of instructions and interactions defining a program and bottlenecks in processing these instructions concurrently. In particular, we note that there is a pending need to improve the message passing between CPU and QPU. This interface raises concerns about the number of qubits and gates as well as matching the clock between these different components. In this contribution, we address some of the considerations for designing new instruction set architectures that may be used to interface the CPU and QPU components, especially as these components become more tightly integrated. Our starting point is the recognition that while single and two-qubit gates may be easiest to implement within hardware, such fine grain description may not be compatible with efficient, large-scale quantum computing.
%%%%%%%%%%%%%%%%%%%%%%%%%%%%%%%%%%%%%%%%%%%%%%%%%%%%%%%%%%%%%%%%%%%%%%%%%%%
\section{RISC and CISC}
RISC architectures conform to a few main principals: segmentation of memory and computational operations, supporting a limited number of basic operations, instruction widths having a firm boundary, maximizing pipelining benefits, and minimizing pipelining penalties \cite{Patterson1985}. These principals give RISC architectures the advantage of standardization and instruction turnover efficiency, but they limit how well RISC architectures can optimize the processing of any particular or complex instruction sequence.
\par
A fixed instruction width is what gives a RISC architecture the ability to pipeline instructions and gives RISC architectures an advantage when compared to CISC architectures for non-anticipated problem classes. As will be illustrated below however, adhering to this fixed instruction width will limit the total usable QPU register size.
\par
Segmenting memory and computation is a given in AQC as the memory is the state of the qubits (and their association to one another) and the computation is the fluctuation of system energy (or alternatively, the passage of time). Segmenting memory and computation in a gate model QPU seems not to have a clear analogy. In the gate model, there is no loading or copying of values into registers between the beginning and ending of the algorithm. Quantum registers (qubits) are simply initialized to a beginning state and then gates are applied to the quantum registers. There are no memory operations other than initialization and reading the final collapsed classical value.
\par
CISC architectures follow the principals of: limited memory registers, emphasizing efficiency improvements through instruction creation and modification, programmer ease, and non-standardized instruction widths \cite{George1990}. CISC architectures become especially attractive in strict domain-oriented processing (like process controllers or vector processors).
\par
In a quantum architecture, the line between memory and computation is inherently blurred as the computation is a process that happens to the memory registers (of which the values cannot be copied), much like as is seen in in-place or in-memory algorithms. However, the CISC concept of optimization through dedicated hardware resources is something that might fit well into a QPU architecture as the use of a QPU is based on a priori knowledge that CPU resources are theoretically inferior to the QPU in terms of processing time or resource efficiency. Thus, a specific set of predefined functions (error correction, quantum Fourier transform, etc.) operating on a specific set of qubits that can be cascaded across other non-preallocated qubits, may be a feature of value in a quantum ISA \cite{Calderbank1996,Hales2000}.
\par
Limiting memory registers, for reasons outlined above, has no corollary in a QPU. Non-standardized instruction widths may serve some purpose if predefined functions are implemented in hardware, but the potential size of that library is not explored in this contribution and only the need for a gate set capable of universal quantum computation is described below. Given these considerations, QPUs and their interfaces to classical controls don't fit neatly into either a RISC or CISC silo as is true for most modern CPUs. However, both ISA models do hold principals that are important in guiding how a QPU ISA should operate. 
%%%%%%%%%%%%%%%%%%%%%%%%%%%%%%%%%%%%%%%%%%%%%%%%%%%%%%%%%%%%%%%%%%%%%%%%%%%
\section{QPU ISA Message Considerations}
In contemplating an ISA for a theoretical gate model QPU (like the IBM Quantum Experience), we draw inspiration from the MIPS ISA J-type (RISC) instruction tuple of $\{opcode, address\}$. From this, we can imagine a similar ISA for quantum hardware of the form $\{opcode, qubit\}$. Taking into account that most circuits will require $m$ qubit gates and that $m$ qubit gates can be reduced to a series of 2-qubit and 1-qubit gates, we expand this quantum ISA to the form $\{opcode, target\_qubit, control\_qubit\}$ \cite{Nielsen2011}. Assuming that the number of elemental 1-qubit and 2-qubit gate types does not exceed 16, our opcode width is then constrained to $2^4$ \cite{Nielsen2011}. Given a classical 64-bit computational and memory architecture, this leaves us 60 bits to specify our qubits. Assuming every qubit is able to participate in any operation as either the target or the control, the target qubit space must be the same width as the control qubit space. Thus, we have 30 bits to specify our target qubit and 30 bits to specify our control. This gives us a hard limit as to how many physical qubits can be used by our system, $2^{30}$ or approximately $1.0737 * 10^9$. Therefore, a 64-bit classical computing architecture limits us to a billion qubit system.
\par
The most obvious solution to this ceiling would be to implement a 128-bit processing and memory ISA, which would expand our addressable qubits to $2^{62}$ or  $4.6117 * 10^{18}$. However, as of 2017, Intel, AMD, and Arch all have publicly stated that there are no current plans to develop a 128-bit processor due to the lack of need. There are possibilities for solutions under the umbrella of a 64-bit architecture. Two qubit gates could be addressed via multiple ISA messages, allowing for $2^{60}$ qubits. This solution would require message management and correlation at the compiler (likely) or even the programmer level (less likely). In addition, this multiple message scheme might require the QPU to adhere to an Execute-Wait-Execute model that would likely reduce the coherence of the qubits in terms of number of gates before decoherence, limiting the depth (length) of quantum circuits.
\par
Additionally, the concept of multiple classical controllers attached to a single QPU is a possibility, in essence dividing the QPU into several different sub-QPUs that could be bridged together when needed \cite{Britt2015}. This scheme would seem to require synchronizing the controllers at the microsecond or even nanosecond level given today's quantum processing technology. In addition, it would seem to imply that there would be specific qubits that are special and exclusive in terms of their spatial relation to the adjoining sub-QPUs and this would complicate instruction compilation, possibly necessitating a need for programmer knowledge of the qubit hardware topology or artificially shrinking the size of quantum registers available for logical computation.
\par
We can draw inspiration from the D-Wave 2000Q concept of QMI when contemplating an ISA for an adiabatic quantum computer. The structure of a QMI instruction is $\{qubit, qubit, value\}$ where if the first and second $qubit$ are equal, the $value$ is the weight to assign to the $qubit$ and if the first and second $qubit$ are not equal, the $value$ is the strength to assign to the coupler between the first $qubit$ and second $qubit$. In addition to the QMI instructions is a header line specifying metadata about how many QMIs are being used. If we strip away the header line (which seems to be an unnecessary construct not at all vital to the adiabatic algorithm), we have what what looks to be a very RISC-like architecture where the instruction widths are uniform and there is a strict segmentation of memory and computation operations. Issues of limited operations types and pipelining really don't fit into the adiabatic model as there is no segmentation of time or function within an adiabatic anneal.
\par
While we don't explore how these instructions are currently passed to the quantum hardware, we theorize that the $\{qubit, qubit, value\}$ is contained in a single message. It has been demonstrated that the current bits of precision (BOP) available in tuning a qubit or coupler in the D-Wave architecture is 10 BOP, which gives us a width for $value$ \cite{Britt2016}. While hardware connectivity constraints don't allow any arbitrary physical qubit to interact with another arbitrary physical qubit, any qubit active in the D-Wave architecture can fill the role of the first or second $qubit$, thus we must assume that the remaining bits of the message are equally divided between the first and second $qubit$. Assuming a 64-bit architecture, each $qubit$ would be allocated 27 bits of width, allowing for $2^{27}$ or $1.34217728 * 10^{8}$ qubits in the system which would allow a fully connected system of approximately 4096 logical qubits if the physical qubits had the same degree of inter-unit-cell connectivity and double the intra-unit-cell connectivity as available in the D-Wave 2000Q \cite{Hamilton2016}.
%%%%%%%%%%%%%%%%%%%%%%%%%%%%%%%%%%%%%%%%%%%%%%%%%%%%%%%%%%%%%%%%%%%%%%%%%%%
\section{Conclusions}
In examining how a QPU might fit into an HPC infrastructure and the necessary interface between classical instructions and quantum processing, we describe potential ISAs for both a gate model QPU and adiabatic model QPU. Our QPU ISAs used a fixed-width message size of 64-bits that if implemented as described would limit the addressable size of a gate model QPU to approximately 1 billion to 4.5 billion billion qubits and the addressable size of an AQC QPU to approximately 134 million qubits. Considerations for logical embedding due to multi-qubit variable types, physical qubit connectivity limitations, and error correction condense the logical qubit work-space in both models.
\par
In trying to create an analogy to classical RISC and CISC architectural features, we find that gate model pipeline processing doesn't seem to fit and a fixed-width message size isn't essential, but might be advantageous in trying to maximize the coherence time of a quantum circuit. Also, issues of segmenting memory from computational tasks are far less of an issue as memory and computation are naturally conjoined (or disjoined depending on perspective) in both the gate model and AQC. The concept of allocating specific quantum register resources to predefined tasks may serve a useful purpose in the gate model (but not likely AQC) as it does in a classical CISC architecture.
\par
Our recommendation is not to orient towards a RISC or CISC architecture when designing future QPU ISAs, but rather we suggest considering the long term consequences of the quantum-classical interface, in particular the message format, on small scale QPUs that might grow into large scale QPUs. Of particular importance is whether the QPU ISA will limit the addressable quantum register size and place an artificial ceiling on QPU scaling.
%%%%%%%%%%%%%%%%%%%%%%%%%%%%%%%%%%%%%%%%%%%%%%%%%%%%%%%%%%%%%%%%%%%%%%%%%%%


\begin{thebibliography}{16}

\bibitem{Grover1996} Grover, Lov K.: A Fast Quantum Mechanical Algorithm for Database Search. STOC '96, 212--219 (1996)

\bibitem{Hen2016} Hen, Itay: Realizable quantum adiabatic search. arXiv:1612.06012 [quant-ph] (2016)

\bibitem{Nielsen2011} Nielsen, Michael A. and Chuang, Isaac L.: Quantum Computation and Quantum Information: 10th Anniversary Edition. Cambridge University Press, New York (2011)

\bibitem{IBM2017} IBM Research Quantum Experience, \url{http://www.research.ibm.com/quantum/}

\bibitem{Farhi2000} E. Farhi, J. Goldstone, S. Gutmann, and M. Sipser: Quantum Computation by Adiabatic Evolution. Report MIT-CTP-2936, Massachusetts Institute of Technology (2000)

\bibitem{DWave2017} The D-Wave 2000Q\texttrademark~System, \url{https://www.dwavesys.com/d-wave-two-system}

\bibitem{Hamilton2016} Kathleen E. Hamilton and Travis S. Humble: Identifying the Minor Set Cover of Dense Connected Bipartite Graphs via Random Matching Edge Sets. 2016, arXiv:1612.07366

\bibitem{Top5002016} Top500.org: Global Supercomputing Capacity Creeps Up as Petascale Systems Blanket Top 100. Top500.org, 2016

\bibitem{Britt2015} Britt, Keith A. and Humble, Travis S.: High-Performance Computing with Quantum Processing Units. J. Emerg. Technol. Comput. Syst. 13, 3, Article 39 (2017)

\bibitem{Fu2016} X. Fu, L. Riesebos, L. Lao, C. G. Almudever, F. Sebastiano, R. Versluis, E. Charbon, and K. Bertels: A heterogeneous quantum computer architecture. Proceedings of the ACM International Conference on Computing Frontiers (CF '16). ACM, New York, NY, USA, 323--330 (2016)

\bibitem{Chuang2004} I. Chuang: qasm2circ, \url{https://www.media.mit.edu/quanta/qasm2circ/}

\bibitem{Patterson1985} Patterson, David A.: Reduced Instruction Set Computers. Commun. ACM, 28, 1, 8--21 (1985)

\bibitem{George1990} A. D. George: An overview of RISC vs. CISC. Proceedings of The Twenty-Second Southeastern Symposium on System Theory, 436--438 (1990)

\bibitem{Calderbank1996} Calderbank, A. R. and Shor, Peter W.: Good quantum error-correction codes exist. Phys. Rev. A, 54, 2, 1098--1105 (1996)

\bibitem{Hales2000} L. Hales and S. Hallgren: An improved quantum Fourier transform algorithm and applications. Proceedings 41st Annual Symposium on Foundations of Computer Science, 5115--525 (2000)

\bibitem{Britt2016} Keith A. Britt and Travis S. Humble: QUBO Computational Reliability via Hamiltonian Engineering. Adiabatic Quantum Computing Conference (2016)

\bibitem{Harrow2009} Aram W. Harrow, Avinatan Hassidim, and Seth Lloyd: Quantum Algorithm for Linear Systems of Equations. Phys. Rev. Lett., 103, 15, 150502 (2009)

\end{thebibliography}
\end{document}